\documentclass[aps,twocolumn]{revtex4}

\usepackage{amsmath}

\usepackage{epsfig}

\def\ov#1{\overline{#1}}

\def\wt#1{\widetilde{#1}}
\def\vb#1{\mbox{\boldmath$#1$}}
\def\pd#1#2{\frac{\partial #1}{\partial #2}}

\def\wh#1{\widehat{#1}}
\def\bdot{\,\vb{\cdot}\,}
\def\btimes{\,\vb{\times}\,}

\def\bhat{\wh{{\sf b}}}

\def\exd{{\sf d}}

\newcommand{\bc}{\begin{center}}
\newcommand{\ec}{\end{center}}
\newcommand{\bt}{\begin{tabbing}}
\newcommand{\et}{\end{tabbing}} 
\newcommand{\be}{\begin{eqnarray*}}
\newcommand{\ee}{\end{eqnarray*}}

\begin{document}

\title{Beyond linear gyrocenter polarization in gyrokinetic theory}

\author{Alain J.~Brizard}
\affiliation{Department of Physics, Saint Michael's College, Colchester, VT 05439, USA}

\begin{abstract}
The concept of polarization in gyrokinetic theory is clarified and generalized to include contributions from the guiding-center (zeroth-order) polarization as well as the nonlinear (second-order) gyrocenter polarization. The guiding-center polarization, which appears as the antecedent (zeroth-order) of the standard linear (first-order) gyrocenter polarization, is obtained from a modified guiding-center transformation. The nonlinear gyrocenter polarization is derived either variationally from the third-order gyrocenter Hamiltonian or directly by gyrocenter push-forward method.
\end{abstract}

\begin{flushright}
September 5, 2013
\end{flushright}

\maketitle

\section{Introduction}

Polarization effects represent one of the hallmarks of nonlinear gyrokinetic theory \cite{Lee_1983,Dubin,Brizard_Hahm} and its numerical implementations 
\cite{Lee_1987,Garbet_2010}. These effects, which are intimately associated with the process of dynamical reduction \cite{Brizard_Vlasovia}, have a long history in plasma physics \cite{Spitzer}. 

Modern gyrokinetic theory \cite{Brizard_Hahm} is based on a two-step phase-space transformation from particle phase space to gyrocenter phase space (with the guiding-center phase-space transformation representing the intermediate step). Since each step generates its own polarization and magnetization effects, the guiding-center polarization and magnetization effects are consequently different from linear (first-order) gyrocenter polarization and magnetization effects. 

While standard nonlinear gyrokinetic theory retains only the first-order (linear) gyrocenter polarization in its gyrokinetic Poisson equation (in the form of a quasineutrality equation used to determine the electrostatic potential through ion polarization effects), it was recently argued by Parra and Catto \cite{Parra_1} that the standard gyrokinetic quasineutrality equation could not be used to find the electrostatic potential in the long-wavelength limit without the introduction of additional nonlinear physics. The ensuing debate \cite{Lee_Koles_1,Parra_2,Lee_Koles_2} concerning the nature of these high-order effects motivates the present work, which hopes to introduce new light into this important topic.

For this purpose, we review in Sec.~\ref{sec:rpm} the derivation of polarization (and magnetization) effects in a general reduced Vlasov-Maxwell theory \cite{Brizard_Vlasovia}. Here, reduced polarization and magnetization can either be derived directly by the push-forward approach or by a variational approach from the reduced Hamiltonian. The geometry of the reduced polarization is also presented. In Sec.~\ref{sec:gc_pol}, we derive the guiding-center polarization and magnetization associated with the guiding-center dynamical reduction \cite{RGL_83,Cary_Brizard}. In the calculation of the guiding-center polarization, however, we find that the standard guiding-center transformation \cite{RGL_83} must be modified in order to recover the standard guiding-center polarization obtained by Pfirsch 
\cite{Pfirsch,Pfirsch_Morrison} and Kaufman \cite{Kaufman_86}. Appendix \ref{sec:gc} summarizes the changes made to the guiding-center transformation in order to recover the Pfirsch-Kaufman guiding-center polarization.

In Sec.~\ref{sec:gy_pol}, we rederive (by push-forward and variational approaches) the standard linear (first-order) gyrocenter polarization that appears in modern gyrokinetic theory \cite{Brizard_Hahm} and derive an expression for the nonlinear (second-order) gyrocenter polarization (derived from the third-order gyrocenter Hamiltonian) that is consistent with the early result of Dubin {\it et al.} \cite{Dubin} and the recent result of Mishchenko and Brizard \cite{Mish_Bri}. We also include a discussion of the recent results of Parra and Catto \cite{Parra_1} and Lee and Kolesnikov \cite{Lee_Koles_1}. We summarize our work in Sec.~\ref{sec:sum} and point out the role played by the guiding-center polarization current density in the formulation of the conservation of toroidal angular momentum in gyrokinetic theory \cite{Brizard_Tronko}.

\section{\label{sec:rpm}Reduced Polarization and Magnetization}

The dynamical reduction of the Vlasov equation, induced by a near-identity transformation on particle phase space
\begin{equation}
{\bf z} \;\rightarrow\; \ov{\bf z} \;\equiv\; {\cal T}_{\epsilon}{\bf z} 
\label{eq:T_epsilon}
\end{equation}
introduces reduced polarization and magnetization effects in the reduced Maxwell equations \cite{Brizard_Vlasovia}:
\begin{eqnarray}
\nabla\bdot{\bf E} & = & 4\pi\,{\varrho} \label{eq:varrho_def} \\
 & \equiv & 4\pi\,\left(\ov{\varrho} \;-\frac{}{} \nabla\bdot\ov{\bf P}\right),
\nonumber \\
\nabla\btimes{\bf B} \;-\; \frac{1}{c}\,\pd{\bf E}{t} & = & \frac{4\pi}{c}\;{\bf J} \label{eq:Jcur_def} \\
 & \equiv & \frac{4\pi}{c}\left(\ov{\bf J} \;+\; \pd{\ov{\bf P}}{t} \;+\; c\,\nabla\btimes\ov{\bf M}\right).
\nonumber
\end{eqnarray}
In Eq.~\eqref{eq:varrho_def}, the particle charge density $\varrho \equiv \ov{\varrho} + \ov{\varrho}_{\rm pol}$ is decomposed in terms of the reduced charge density 
\begin{equation}
\ov{\varrho} \;=\; e\,\int\,\ov{F}\,d^{3}\ov{p}, 
\label{eq:rho_bar_def}
\end{equation}
where $\ov{F}$ denotes the reduced Vlasov distribution (summation over particle species is henceforth implied) and the polarization charge density 
$\ov{\varrho}_{\rm pol} \equiv -\,\nabla\bdot\ov{\bf P}$, which is defined in terms of the reduced polarization $\ov{\bf P}$. In Eq.~\eqref{eq:Jcur_def}, the particle current density ${\bf J} \equiv \ov{\bf J} + \ov{\bf J}_{\rm pol} + \ov{\bf J}_{\rm mag}$ is decomposed in terms of the reduced current density 
\begin{equation}
\ov{\bf J} \;=\; e\,\int\,\ov{F}\,\frac{d_{\epsilon}\ov{\bf x}}{dt}\,d^{3}\ov{p}, 
\label{eq:J_bar_def}
\end{equation}
where $d_{\epsilon}\ov{\bf x}/dt$ denotes the reduced particle velocity, the polarization current density $\ov{\bf J}_{\rm pol} \equiv 
\partial\ov{\bf P}/\partial t$, and the magnetization current $\ov{\bf J}_{\rm mag} \equiv c\,\nabla\btimes\ov{\bf M}$, which is defined in terms of the reduced magnetization $\ov{\bf M}$. Here, the reduced Vlasov distribution $\ov{F}$ satisfies the reduced Vlasov equation
\begin{equation}
0 \;=\; \pd{\ov{F}}{t} \;+\; \frac{d_{\epsilon}\ov{z}^{\alpha}}{dt}\;\pd{\ov{F}}{\ov{z}^{\alpha}} \;\equiv\; \pd{\ov{F}}{t} \;+\; \{ \ov{F},\; \ov{H}
\}_{\epsilon},
\label{eq:ov_Vlasov}
\end{equation}
where the reduced Hamiltonian $\ov{H}$ and the reduced Poisson bracket $\{\;,\;\}_{\epsilon}$ are both derived by Lie-transform perturbation methods 
\cite{Brizard_Vlasovia}. Since the process of dynamical reduction has eliminated a specific fast time scale from the reduced particle dynamics 
$d_{\epsilon}\ov{z}^{\alpha}/dt$, the reduced Vlasov distribution $\ov{F}$ is a constant on this fast time scale.

The reduced polarization $\ov{\bf P}$ and the reduced magnetization $\ov{\bf M}$ appearing in the reduced Maxwell equations 
\eqref{eq:varrho_def}-\eqref{eq:Jcur_def} can be derived by two different approaches: a ``bottom-up'' push-forward approach that builds directly on the near-identity transformation \eqref{eq:T_epsilon}; or a ``top-down'' approach based on the explicit dependence of the reduced Hamiltonian $\ov{H}$ on the electric and magnetic fields, respectively. 

\subsection{Bottom-up push-forward approach}

The reduced polarization $\ov{\bf P}$ and magnetization $\ov{\bf M}$ can be constructed directly by push-forward method from the reduced displacement 
\cite{Brizard_Vlasovia} defined as the difference between the push-forward ${\sf T}_{\epsilon}^{-1}{\bf x}$ of the particle position ${\bf x}$ and the reduced particle position $\ov{\bf x}$:
\begin{eqnarray}
\vb{\rho}_{\epsilon} & \equiv & {\sf T}_{\epsilon}^{-1}{\bf x} \;-\; \ov{\bf x} \nonumber \\
 & = & -\,\epsilon\;G_{1}^{{\bf x}} \,-\, \epsilon^{2} \left( G_{2}^{{\bf x}} - \frac{1}{2}\,{\sf G}_{1}\cdot\exd G_{1}^{{\bf x}} \right) + \cdots,
\label{eq:rho_epsilon_def}
\end{eqnarray}
where the generating vector fields $({\sf G}_{1}, {\sf G}_{2}, \cdots)$ are associated with the phase-space transformation \eqref{eq:T_epsilon}. 

For the reduced polarization, we begin with the push-forward of the particle charge density 
\begin{eqnarray}
\varrho & = & e\;\int\,f\,d^{3}p \;=\; e\;\int\,f\,\delta^{3}({\bf x} - {\bf r})\,d^{6}z \nonumber \\
 & = & e\;\int\;\ov{F}\;\left\langle\delta^{3}(\ov{\bf x} + \vb{\rho}_{\epsilon} - {\bf r})\right\rangle\;d^{6}\ov{z},
\label{eq:rho_def}
\end{eqnarray}
where the particle charge density is first evaluated as an integral over the entire particle phase space (with only particles whose positions ${\bf x}$ are at the field point ${\bf r}$ contributing to $\varrho$) and then transformed to reduced phase space. The reduced phase-space integral now involves the averaged push-forward delta function $\langle{\sf T}_{\epsilon}^{-1}\delta^{3}({\bf x} - {\bf r})\rangle \equiv \langle\delta^{3}(\ov{\bf x} + 
\vb{\rho}_{\epsilon} - {\bf r})\rangle$, where $\langle\cdots\rangle$ denotes averaging over the fast orbital time scale, and the reduced Vlasov distribution $\ov{F} \equiv {\sf T}_{\epsilon}^{-1}f$ is the push-forward of the particle Vlasov distribution $f$. 

When Eq.~\eqref{eq:rho_def} is Taylor-expanded as a multipole expansion in powers of $\vb{\rho}_{\epsilon}$ (e.g., dipole + quadrupole + $\cdots$), we obtain the expression
\begin{equation}
\varrho \;\equiv\; \ov{\varrho} \;-\; \nabla\bdot\ov{\bf P},
\label{eq:rho_particle}
\end{equation}
where the reduced polarization \cite{Brizard_Vlasovia} associated with the transformation \eqref{eq:T_epsilon} is defined as
\begin{eqnarray}
\ov{{\bf P}} & \equiv & \int \left[ \vb{\pi}_{\epsilon}\,\ov{F} - \left(\frac{e}{2}\,\langle\vb{\rho}_{\epsilon}\,
\vb{\rho}_{\epsilon}\rangle\bdot\nabla\ov{F} + \cdots\right)\right]\,d^{3}\ov{p}.
\label{eq:pol_def}
\end{eqnarray}
Here, the reduced polarization density
\begin{equation}
\vb{\pi}_{\epsilon} \;\equiv\; e\,\left(\langle\vb{\rho}_{\epsilon}\rangle \;-\; \nabla\bdot\left\langle\frac{1}{2}\;\vb{\rho}_{\epsilon}\,
\vb{\rho}_{\epsilon}\right\rangle \;+\; \cdots\right)
\label{eq:pi_epsilon}
\end{equation}
includes contributions from the electric dipole and quadrupole moments (higher-order multipole contributions are ignored in the present work), while the remaining terms in Eq.~\eqref{eq:pol_def} represent finite-Larmor-radius (FLR) corrections to the reduced Vlasov distribution. 

\begin{figure}
\epsfysize=1.25in
\epsfbox{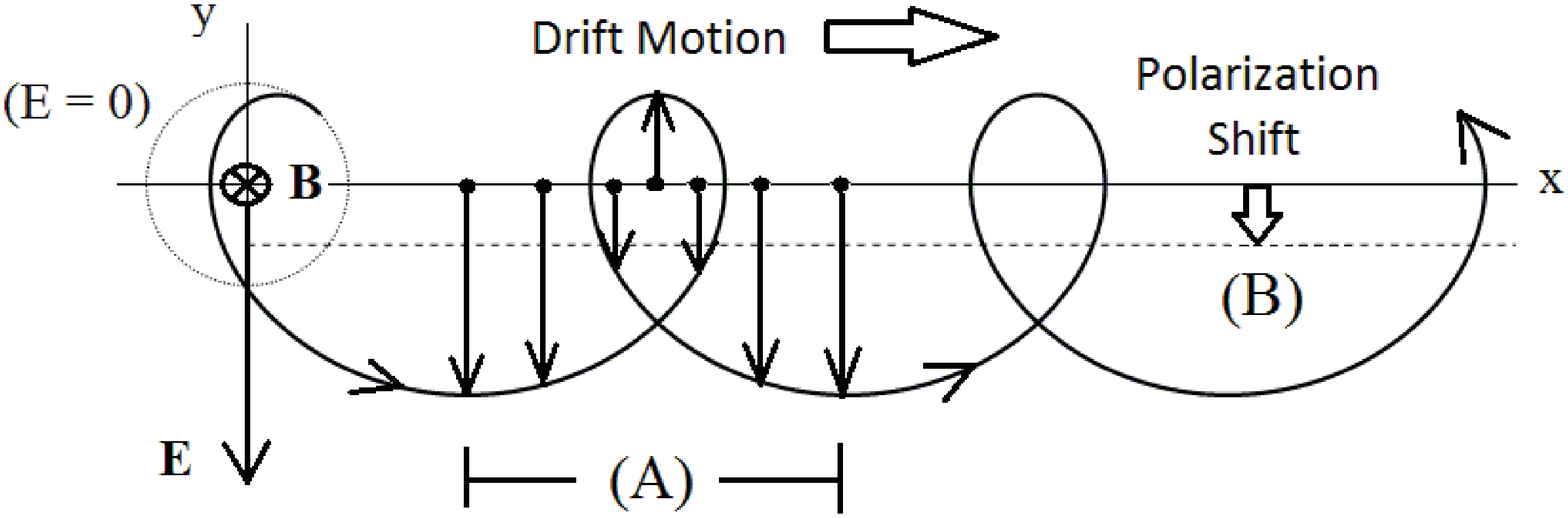}
\caption{The drifting motion (dark curve) of a gyrating charged particle in crossed electric and magnetic fields leads to a polarization shift from the guiding-center position (on the $x$-axis) to the averaged particle position (on the dotted horizontal line). In the absence of drift motion 
$({\bf E} = 0)$, the particle's gyromotion is along a circular orbit (centered at the origin).}
\label{fig:Drift_pol}
\end{figure}

The geometry of the reduced polarization density \eqref{eq:pi_epsilon} is shown in Fig.~\ref{fig:Drift_pol} for the case of a charged particle moving in a strong magnetic field ${\bf B} = -\,B\,\wh{\sf z}$ (into the page) in the presence of a uniform electric field ${\bf E} = -\,E\,\wh{\sf y}$, which yields a drifting particle orbit (dark curve) that is moving to the right (along the $x$-axis) with $E\times B$ drift velocity ${\bf E}\btimes c\,\bhat/B = (c\,E/B)\,\wh{\sf x}$. In part (A) of Fig.~\ref{fig:Drift_pol}, the up-and-down vertical arrows show the reduced displacement 
\eqref{eq:rho_epsilon_def}, defined as the difference between the particle position (on the dark curve) and the reduced position (on the $x$-axis), at various points during a fast gyration period. Part (B) shows the polarization {\it shift} (from the $x$-axis to the dotted horizontal line), which is  perpendicular to both the magnetic-field direction and the drift velocity. 

For a general drift-particle orbit in a strong magnetic field, the reduced polarization density \eqref{eq:pi_epsilon} is defined to lowest order 
\cite{Spitzer} as 
\begin{equation}
\vb{\pi}_{\epsilon} \;\equiv\; \frac{e\,\bhat}{\Omega}\btimes\frac{d_{\epsilon}\ov{\bf x}}{dt},
\label{eq:Pi_Spitzer}
\end{equation}
where $\Omega = e\,B/(mc)$ denotes the gyrofrequency for a particle of charge $e$ and mass $m$. We note that the ratio of the magnitude of the polarization shift (defined as $|\vb{\pi}_{\epsilon}|/e$) to the gyroradius $\rho_{\bot}$ of the undrifting particle (i.e., the circle's radius in 
Fig.~\ref{fig:Drift_pol}) can be used as the definition of the dimensionless parameter $\epsilon \equiv |\vb{\pi}_{\epsilon}|/(e\,\rho_{\bot}) \ll 1$ (the polarization shift is exaggerated in Fig.~\ref{fig:Drift_pol} for illustrative purposes).

For the reduced magnetization, we consider the push-forward of the particle current density
\begin{eqnarray}
{\bf J} & = & e\;\int\,f\,\frac{d{\bf x}}{dt}\;d^{3}p \;=\; e\;\int\,f\,\frac{d{\bf x}}{dt}\,\delta^{3}({\bf x} - {\bf r})\;d^{6}z 
\label{eq:J_particle} \\
 & = & e\;\int\;\ov{F}\;\left\langle \left(\frac{d_{\epsilon}\ov{\bf x}}{dt} + \frac{d_{\epsilon}\vb{\rho}_{\epsilon}}{dt} \right)
\delta^{3}(\ov{\bf x} + \vb{\rho}_{\epsilon} - {\bf r})\right\rangle d^{6}\ov{z},
\nonumber
\end{eqnarray}
where the push-forward of the particle velocity 
\[ {\sf T}_{\epsilon}^{-1}\left(\frac{d{\bf x}}{dt}\right) \;\equiv\; \frac{d_{\epsilon}\ov{\bf x}}{dt} \;+\; 
\frac{d_{\epsilon}\vb{\rho}_{\epsilon}}{dt} \]
is decomposed into the reduced particle velocity $d_{\epsilon}\ov{\bf x}/dt$ and the reduced displacement velocity $d_{\epsilon}\vb{\rho}_{\epsilon}/dt$. A multipole expansion of Eq.~\eqref{eq:J_particle} yields the expression for the plasma current density
\begin{equation}
{\bf J} \;\equiv\; \ov{\bf J} \;+\; \pd{\ov{\bf P}}{t} \;+\; c\;\nabla\btimes\ov{\bf M},
\label{eq:J_def}
\end{equation}
where, using the expression \eqref{eq:pol_def} for the reduced polarization, the reduced polarization current in Eq.~\eqref{eq:J_def} is first expressed 
as
\begin{eqnarray*}
\pd{\ov{\bf P}}{t} & = & \int d^{3}\ov{p} \left[e\,\left( \ov{F}\;\pd{\langle\vb{\rho}_{\epsilon}\rangle}{t} \;+\; \langle\vb{\rho}_{\epsilon}\rangle\,
\pd{\ov{F}}{t} \right) \right. \\
 &  &\left.-\, \frac{e}{2}\;\nabla\bdot\left( \langle\vb{\rho}_{\epsilon}\,\vb{\rho}_{\epsilon}\rangle\;\pd{\ov{F}}{t} \;+\; \ov{F}\;\pd{}{t}\left\langle
\vb{\rho}_{\epsilon}\,\vb{\rho}_{\epsilon}\right\rangle \;+\; \cdots \right)\right]. 
\end{eqnarray*}
When the reduced Vlasov equation \eqref{eq:ov_Vlasov} is used to replace $\partial\ov{F}/\partial t$, we obtain
\begin{widetext}
\begin{equation}
\pd{\ov{\bf P}}{t} \;=\; e\;\int\;\left(\frac{d_{\epsilon}\langle\vb{\rho}_{\epsilon}\rangle}{dt}\right)\;\ov{F}\,d^{3}\ov{p} \;-\;\nabla\bdot\left[ e\,\int \left( \frac{d_{\epsilon}\ov{\bf x}}{dt}\;\langle\vb{\rho}_{\epsilon}\rangle \;+\; \frac{1}{2}\,\left\langle
\frac{d_{\epsilon}\vb{\rho}_{\epsilon}}{dt}\,\vb{\rho}_{\epsilon} + \vb{\rho}_{\epsilon}\,\frac{d_{\epsilon}\vb{\rho}_{\epsilon}}{dt}\right\rangle\right)\ov{F}\,d^{3}\ov{p} \;+\; \cdots \right],
\label{eq:pol_current}
\end{equation}
\end{widetext}
i.e., the reduced polarization current is defined, to lowest order, as the reduced Vlasov-moment of the averaged reduced displacement velocity 
$d_{\epsilon}\langle\vb{\rho}_{\epsilon}\rangle/dt$. If we now insert Eq.~\eqref{eq:pol_current} into Eq.~\eqref{eq:J_particle}, we obtain the reduced magnetization current
\begin{eqnarray}
c\,\nabla\btimes\ov{\bf M} & \equiv & {\bf J} \;-\; \left(\ov{\bf J} \;+\; \pd{\ov{\bf P}}{t} \right) \\
 & = & \nabla\bdot\left[ e\,\int \left(\frac{d_{\epsilon}\ov{\bf x}}{dt}\;\langle\vb{\rho}_{\epsilon}\rangle
- \langle\vb{\rho}_{\epsilon}\rangle\;\frac{d_{\epsilon}\ov{\bf x}}{dt}\right) \ov{F}\,d^{3}\ov{p} \right. \nonumber \\
 &  &\left.+\, \frac{e}{2} \int \left\langle\frac{d_{\epsilon}\vb{\rho}_{\epsilon}}{dt}\,\vb{\rho}_{\epsilon}
- \vb{\rho}_{\epsilon}\,\frac{d_{\epsilon}\vb{\rho}_{\epsilon}}{dt}\right\rangle \ov{F}\,d^{3}\ov{p}\right]. \nonumber
\end{eqnarray}
By using the vector identity $\nabla\bdot({\bf K}\,{\bf G} - {\bf G}\,{\bf K}) = \nabla\btimes({\bf G}\btimes{\bf K})$, which is valid for arbitrary vector fields $({\bf G}, {\bf K})$, we obtain the reduced (dipole) magnetization \cite{Brizard_Vlasovia}
\begin{equation}
\ov{{\bf M}} \equiv \frac{e}{c}\int\ov{F}\,\left(\frac{1}{2}\,\left\langle\vb{\rho}_{\epsilon}\btimes\frac{d_{\epsilon}\vb{\rho}_{\epsilon}}{dt}\right\rangle + \langle\vb{\rho}_{\epsilon}\rangle\btimes\frac{d_{\epsilon}\ov{\bf x}}{dt} \right) d^{3}\ov{p} + \cdots.
\label{eq:mag_def}
\end{equation}
Here, the reduced intrinsic magnetization density is defined as
\begin{equation}
\vb{\mu}_{\epsilon} \;\equiv\; \frac{e}{2c}\;\left\langle\vb{\rho}_{\epsilon}\btimes \frac{d_{\epsilon}\vb{\rho}_{\epsilon}}{dt}\right\rangle
\label{eq:mu_epsilon}
\end{equation}
while 
\begin{equation}
\frac{e}{c}\,\langle\vb{\rho}_{\epsilon}\rangle\btimes \frac{d_{\epsilon}\ov{\bf x}}{dt} \;\simeq\; \vb{\pi}_{\epsilon}\btimes \frac{1}{c}\,
\frac{d_{\epsilon}\ov{\bf x}}{dt}
\label{eq:moving_epsilon}
\end{equation}
represents the moving-electric-dipole contribution \cite{Jackson}. 

For each near-identity phase-space transformation \eqref{eq:T_epsilon}, a reduced displacement \eqref{eq:rho_epsilon_def} can, therefore, be constructed from which a reduced polarization \eqref{eq:pol_def} and a reduced magnetization \eqref{eq:mag_def} can be derived.
 
\subsection{Top-down variational approach}

As a result of the near-identity transformation \eqref{eq:T_epsilon}, the reduced Hamiltonian $\ov{H}$ appearing in the reduced Vlasov equation 
\eqref{eq:ov_Vlasov} acquires a dependence on the electromagnetic fields $({\bf E}, {\bf B})$. The reduced polarization density \eqref{eq:pi_epsilon} and the reduced intrinsic magnetization density \eqref{eq:mu_epsilon} can thus be derived from the reduced Hamiltonian as follows.

From a variational point of view, the reduced polarization density \eqref{eq:pi_epsilon} is defined from the reduced Hamiltonian $\ov{H}$ as
\begin{equation}
\vb{\pi}_{\epsilon} \;\equiv\; -\;\pd{\ov{H}}{\bf E} \;+\; \nabla\bdot\left(\pd{\ov{H}}{\nabla{\bf E}}\right) \;+\; \cdots, 
\label{eq:pi_epsilon_var}
\end{equation}
where the first and second terms represent the electric dipole and quadrupole contributions, respectively. The reduced intrinsic magnetization density \eqref{eq:mu_epsilon}, on the other hand, is defined as
\begin{equation}
\vb{\mu}_{\epsilon} \;\equiv\; -\;\pd{\ov{H}}{\bf B} \;+\; \pd{\ov{H}}{\bf E}\btimes\frac{1}{c}\,\frac{d_{\epsilon}\ov{\bf x}}{dt}.
\label{eq:mu_epsilon_var}
\end{equation}
We now relate these variational definitions with the push-forward definitions \eqref{eq:pi_epsilon} and \eqref{eq:mu_epsilon} as follows.

We begin our variational derivation of Eqs.~\eqref{eq:pi_epsilon_var}-\eqref{eq:mu_epsilon_var} by considering the ``interaction'' Hamiltonian
\begin{equation}
h \;\equiv\; e\,\Phi({\bf x},t) \;-\; \frac{e}{c}\,\frac{d{\bf x}}{dt}\bdot{\bf A}({\bf x},t),
\label{eq:H_EM}
\end{equation}
which is expressed in terms of the electromagnetic potentials $(\Phi, {\bf A})$. We note that the Hamiltonian \eqref{eq:H_EM} transforms as
$h \rightarrow h - (e/c)\,d\chi/dt$ under the gauge transformation $(\Phi, {\bf A}) \rightarrow (\phi - c^{-1}\partial\chi/\partial t, {\bf A} + \nabla\chi)$; hence, an arbitrary exact time derivative can be added to the Hamiltonian \eqref{eq:H_EM} without changing the Hamiltonian dynamics. 

First, we consider the push-forward of the Hamiltonian \eqref{eq:H_EM} induced by the phase-space transformation \eqref{eq:T_epsilon}:
\begin{eqnarray}
{\sf T}_{\epsilon}^{-1}\,h & = & e\,\Phi(\ov{\bf x} + \vb{\rho}_{\epsilon}) - \frac{e}{c}\left(\frac{d_{\epsilon}\ov{\bf x}}{dt} + 
\frac{d_{\epsilon}\vb{\rho}_{\epsilon}}{dt}\right)\bdot{\bf A}(\ov{\bf x} + \vb{\rho}_{\epsilon}) \nonumber \\
 & \equiv & \ov{H} \;+\;  \frac{d_{\epsilon}\sigma}{dt}, 
\label{eq:delta_H}
\end{eqnarray}
where the gauge term $d_{\epsilon}\sigma/dt$ is introduced to simplify $\ov{H}$ and eliminate the fast-time-scale dependence of ${\sf T}_{\epsilon}^{-1}\,h$. Next, we Taylor-expand the electromagnetic potentials in Eq.~\eqref{eq:delta_H} up to second (quadrupole) order $\vb{\rho}_{\epsilon}$ and obtain the fast-time-averaged expression
\begin{eqnarray}
\ov{H} & = & e \left( \ov{\Phi} \;-\; \frac{1}{c}\,\ov{\bf A}\bdot\frac{d_{\epsilon}\ov{\bf x}}{dt} \right) \label{eq:delta_H_dipole} \\
 &  &-\; e\left( \langle\vb{\rho}_{\epsilon}\rangle\bdot\ov{\bf E} \;+\; \frac{1}{2}\,\langle\vb{\rho}_{\epsilon}\,\vb{\rho}_{\epsilon}\rangle\,:\,\
\ov{\nabla}\ov{\bf E}\right) \nonumber \\
 &  &-\; \frac{e}{c}\,\left(\langle\vb{\rho}_{\epsilon}\rangle\btimes\frac{d_{\epsilon}\ov{\bf x}}{dt} \;+\; \frac{1}{2}\,\left\langle
\vb{\rho}_{\epsilon}\btimes\frac{d_{\epsilon}\vb{\rho}_{\epsilon}}{dt}\right\rangle\right)\bdot\ov{\bf B},
\nonumber
\end{eqnarray}
where the perturbed electromagnetic fields are evaluated at the reduced position $\ov{\bf x}$. We now use the definition \eqref{eq:pi_epsilon_var} to obtain the reduced polarization density \eqref{eq:pi_epsilon}:
\begin{eqnarray}
\vb{\pi}_{\epsilon} & = & -\;\pd{\ov{H}}{\ov{\bf E}} \;+\; \ov{\nabla}\bdot\left(\pd{\ov{H}}{\ov{\nabla}\ov{\bf E}}\right) \nonumber \\
 & = & e\,\langle\vb{\rho}_{\epsilon}\rangle
\;-\; \frac{e}{2}\;\ov{\nabla}\bdot\langle\vb{\rho}_{\epsilon}\,\vb{\rho}_{\epsilon}\rangle, 
\label{eq:H_E_dipole}
\end{eqnarray}
and the definition \eqref{eq:mu_epsilon_var} to obtain the reduced magnetization density \eqref{eq:mu_epsilon}:
\begin{eqnarray}
\vb{\mu}_{\epsilon} & = & -\;\pd{\ov{H}}{\ov{\bf B}} \;+\; \pd{\ov{H}}{\ov{\bf E}}\btimes\frac{1}{c}\,\frac{d_{\epsilon}\ov{\bf x}}{dt} \nonumber \\
 & = & \frac{e}{2c}\,\left\langle\vb{\rho}_{\epsilon}\btimes\frac{d_{\epsilon}\vb{\rho}_{\epsilon}}{dt}\right\rangle.
\label{eq:H_B_dipole}
\end{eqnarray}

\subsection{Oscillation-center polarization \& magnetization}

We illustrate the concepts of reduced polarization and magnetization by briefly reviewing the polarization and magnetization that arise following the oscillation-center transformation \cite{Cary_Kaufman,Brizard_JPCS}. First, we consider the eikonal representation of the high-frequency fluctuating fields
\begin{equation} 
(\wt{\bf E}_{1},\wt{\bf B}_{1}) \;=\; \left({\bf E}_{1},\frac{}{} {\bf B}_{1}\right)\;\exp\left(i\,\epsilon^{-1}\Theta(\epsilon{\bf r},\epsilon t)\right) \;+\; {\rm c.c.},
\label{eq:eikonal_EB}
\end{equation}
where the eikonal phase $\Theta$ yields the ``local'' definitions of wave-frequency $\omega \equiv -\,\epsilon^{-1}\partial_{t}\Theta$ and wave-vector
${\bf k} \equiv \epsilon^{-1}\nabla\Theta$, the complex-valued eikonal amplitudes $({\bf E}_{1},{\bf B}_{1})$ are weak functions of space and time, and the eikonal parameter $\epsilon \ll 1$ is associated with the fluctuating-field amplitude. The Lie-transform analysis leading to the oscillation-center dynamics \cite{Cary_Kaufman} (which is independent of the fast wave space-time scales) yields the first-order oscillation-center displacement 
$\wt{\vb{\xi}}_{1} \equiv \vb{\xi}_{1{\rm oc}}\,\exp(i\epsilon^{-1}\Theta) + {\rm c.c.}$, where the complex-valued eikonal amplitude 
$\vb{\xi}_{1{\rm oc}}$ is
\begin{equation}
\vb{\xi}_{1{\rm oc}} \;\equiv\; \frac{-\,e}{m\,\omega^{\prime 2}} \left( {\bf E}_{1} \;+\; \frac{\bf v}{c}\btimes{\bf B}_{1}\right),
\label{eq:xi_oc}
\end{equation}
and $\omega^{\prime} \equiv \omega - {\bf k}\bdot{\bf v}$ denotes the Doppler-shifted wave frequency. In the present case, fast-time-scale averaging (denoted as $\langle\cdots\rangle_{\rm oc}$) involves an average over the eikonal phase $\Theta$. We note here that, because of the mass scaling 
$m^{-1}$ in Eq.~\eqref{eq:xi_oc}, the electron oscillation-center displacement is much larger than the ion oscillation-center displacement by a factor of the ion-over-electron mass ratio.

In Ref.~\cite{Brizard_JPCS}, the oscillation-center polarization and magnetization are derived in the limit of small background electric and magnetic fields $({\bf E}_{0}, {\bf B}_{0})$, where the second-order oscillation-center Hamiltonian is 
\begin{eqnarray}
H_{2{\rm oc}} & = & m\,\omega^{\prime 2}\;|{\vb{\xi}}_{1{\rm oc}}|^{2} \;-\; {\bf E}_{0}\bdot\vb{\pi}_{2{\rm oc}} \nonumber \\
 &  &-\; {\bf B}_{0}\bdot\left( \vb{\mu}_{2{\rm oc}} \;+\; \vb{\pi}_{2{\rm oc}}\btimes\frac{{\bf v}}{c}\right).
\label{eq:H2_oc}
\end{eqnarray} 
Here, the lowest-order term 
\[ \frac{m}{2}\,\left\langle\left|\frac{d\wt{\vb{\xi}}_{1}}{dt}\right|\right\rangle_{\rm oc} \;=\; m\,\omega^{\prime 2}\;|{\vb{\xi}}_{1{\rm oc}}|^{2} \] 
represents the standard oscillation-center ponderomotive Hamiltonian \cite{Cary_Kaufman}, while the oscillation-center electric-dipole moment 
$\vb{\pi}_{2{\rm oc}}$ and the intrinsic magnetic-dipole moment $\vb{\mu}_{2{\rm oc}}$ are expressed in terms of the oscillation-center displacement \eqref{eq:xi_oc} as
\begin{eqnarray} 
\vb{\pi}_{2{\rm oc}} & \equiv & -\,\pd{H_{2{\rm oc}}}{{\bf E}_{0}} \;=\; e\,{\bf k}\btimes\left(i\,\vb{\xi}_{1{\rm oc}}\btimes
\vb{\xi}_{1{\rm oc}}^{*}\right), \label{eq:pi_oc} \\
\vb{\mu}_{2{\rm oc}} & \equiv & -\,\pd{H_{2{\rm oc}}}{{\bf B}_{0}} \;+\; \pd{H_{2{\rm oc}}}{{\bf E}_{0}}\btimes\frac{\bf v}{c} \nonumber \\
 & = & \frac{e}{c}\,\omega^{\prime}\left(i\,\vb{\xi}_{1{\rm oc}}\btimes\vb{\xi}_{1{\rm oc}}^{*}\right). 
\label{eq:mu_oc}
\end{eqnarray}
These expressions can also be obtained directly by push-forward method from the oscillation-center displacement
\begin{eqnarray}
\vb{\rho}_{\rm oc} & = & \epsilon\,\wt{\vb{\xi}}_{1} \;+\; \frac{\epsilon^{2}}{2} \left( \wt{\vb{\xi}}_{1}\bdot\nabla\wt{\vb{\xi}}_{1} \;+\;
\frac{d\wt{\vb{\xi}}_{1}}{dt}\bdot\pd{\wt{\vb{\xi}}_{1{\rm oc}}}{\bf v} \right) \nonumber \\
 &  &-\; \epsilon^{2}\,G_{2}^{\bf x} \;+\; \cdots,
\label{eq:rho_oc}
\end{eqnarray}
where $G_{1}^{\bf x} = -\,\wt{\vb{\xi}}_{1}$, $G_{1}^{\bf v} = d\wt{\vb{\xi}}_{1}/dt$, and the second-order field $G_{2}^{\bf x}$ is not needed. The eikonal-average of the oscillation-center displacement \eqref{eq:rho_oc} yields the second-order averaged displacement
\begin{eqnarray}
\langle\vb{\rho}_{\rm oc}\rangle_{\rm oc} & = & \epsilon^{2}\;{\rm Re}\left[ i\,{\bf k}\bdot\left(\vb{\xi}_{1{\rm oc}}^{*}\frac{}{}
\vb{\xi}_{1{\rm oc}}\right) \;-\; i\,\omega^{\prime}\;\vb{\xi}_{1{\rm oc}}\bdot\pd{\vb{\xi}_{1{\rm oc}}^{*}}{\bf v} \right] \nonumber \\
 & = & \epsilon^{2}\;{\bf k}\btimes\left(i\,\vb{\xi}_{1{\rm oc}}\btimes\vb{\xi}_{1{\rm oc}}^{*}\right),
\label{eq:rho_oc_ave}
\end{eqnarray}
from which we recover Eq.~\eqref{eq:pi_oc}. Using Eq.~\eqref{eq:mu_epsilon}, we obtain the eikonal average
\begin{equation}
\frac{1}{2}\;\left\langle\vb{\rho}_{\rm oc}\btimes\frac{d\vb{\rho}_{\rm oc}}{dt}\right\rangle_{\rm oc} \;=\; \epsilon^{2}\,\omega^{\prime}\;\left(i\,
\wt{\vb{\xi}}_{1{\rm oc}}\btimes\wt{\vb{\xi}}_{1{\rm oc}}^{*}\right),
\label{eq:mu_oc_ave}
\end{equation}
from which we recover Eq.~\eqref{eq:mu_oc}.

We will show in Sec.~\ref{sec:gy_pol} [see Eq.~\eqref{eq:pi2_gy_eik}] how Eq.~\eqref{eq:pi_oc} reveals some universal features of reduced polarization as a result of the process of dynamical reduction.

\section{\label{sec:gc_pol}Guiding-center Polarization and Magnetization}

We now proceed with the process of dynamical reduction associated with gyrokinetic theory, whereby the fast gyromotion perpendicular to a magnetic-field line is decoupled from the slow parallel motion along and drift motion across magnetic-field lines (see Fig.~\ref{fig:Drift_pol}). Standard gyrokinetic theory \cite{Brizard_Hahm} is based on a two-step near-identity transformation ${\cal T}_{\epsilon} \equiv {\cal T}_{\rm gy}\,{\cal T}_{\rm gc}$ starting with the guiding-center transformation (from particle phase space) followed by the gyrocenter transformation (from guiding-center phase space).

In the present Section, we derive the guiding-center polarization and magnetization densities \eqref{eq:pi_epsilon} and \eqref{eq:mu_epsilon} [or 
\eqref{eq:pi_epsilon_var} and \eqref{eq:mu_epsilon_var}] associated with the reduced guiding-center displacement 
\begin{equation}
\vb{\rho}_{\rm gc} \;=\; \vb{\rho}_{0} \;+\; \epsilon_{\rm B}\,\vb{\rho}_{1{\rm gc}} \;+\; \cdots, 
\label{eq:rho_gc}
\end{equation}
where $\vb{\rho}_{0} = -\,G_{1}^{\bf x} \equiv (2\mu B/m\Omega^{2})^{1/2}\wh{\rho}$ denotes the gyroangle-dependent lowest-order gyroradius (whose gyroangle average $\langle\vb{\rho}_{0}\rangle \equiv 0$ vanishes identically) and the first-order correction $\vb{\rho}_{1{\rm gc}}$ is presented in 
App.~\ref{sec:gc} ($\epsilon_{\rm B}$ denotes the small parameter associated with the nonuniformity of the background magnetic field). In the next Section, we will derive the linear (first-order) and nonlinear (second-order) gyrocenter corrections to the reduced polarization due to the gyrocenter phase-space transformation.

\subsection{Guiding-center polarization}

Using Eq.~\eqref{eq:pi_epsilon}, we derive the gyroangle-averaged guiding-center polarization density $\vb{\pi}_{\rm gc} \equiv \epsilon_{\rm B}\,
\vb{\pi}_{1{\rm gc}} + \cdots$, where the lowest-order guiding-center polarization density is given by the lowest-order Pfirsch-Kaufman expression 
\cite{Pfirsch,Pfirsch_Morrison,Kaufman_86}
\begin{eqnarray}
\vb{\pi}_{1{\rm gc}} & \equiv & e\,\langle\vb{\rho}_{1{\rm gc}}\rangle - \nabla\bdot\left( \frac{e}{2}\,\left\langle\vb{\rho}_{0}\frac{}{}\vb{\rho}_{0}\right\rangle \right) \nonumber \\
 & = & \frac{e\,\bhat}{\Omega}\btimes \frac{d_{\rm gc}{\bf X}}{dt},
\label{eq:pi_gc}
\end{eqnarray}
with the perpendicular guiding-center drift-velocity defined as
\begin{equation}
\left(\frac{d_{\rm gc}{\bf X}}{dt}\right)_{\bot} \;\equiv\; \frac{\bhat}{m\Omega}\btimes\left( \mu\,\nabla B \;+\; \frac{p_{\|}^{2}}{m}\,\bhat\bdot\nabla\bhat\right).
\label{eq:gc_velocity}
\end{equation} 
Details of the calculation leading to Eq.~\eqref{eq:pi_gc} are given in App.~\ref{sec:gc}. Using Eq.~\eqref{eq:pol_def}, the guiding-center polarization
\begin{equation}
{\bf P}_{\rm gc} \;\equiv\; \int\left[ F\,\left( \frac{e\,\bhat}{\Omega}\btimes\frac{d_{\rm gc}{\bf X}}{dt}\right) \;-\; \frac{\mu\,B}{2\,m\Omega^{2}}\;
\nabla_{\bot}F \right] d^{3}P
\label{eq:P_gc_final}
\end{equation}
is, therefore, a first-order term in magnetic-field (and Vlasov) nonuniformity. 

We note that, because of the presence of a spatial divergence in its definition, the guiding-center polarization \eqref{eq:P_gc_final} yields a guiding-center polarization charge density $\varrho_{\rm pol\,(gc)}^{(2)} \equiv -\,\nabla\bdot{\bf P}_{\rm gc}^{(1)}$ that is ordered at second order in magnetic-field (and Vlasov) nonuniformity, which explains why it is often ignored in the gyrokinetic Poisson equation \cite{Brizard_Hahm}. The second-order polarization current density ${\bf J}_{\rm pol\,(gc)}^{(2)} \equiv \partial{\bf P}_{\rm gc}^{(1)}/\partial t$, however, is of the same order as the gyrocenter polarization current density and, therefore, cannot be ignored. The guiding-center polarization current density derived from 
Eq.~\eqref{eq:P_gc_final} has, in fact, recently appeared in the derivation of momentum conservation laws associated with nonlinear gyrokinetic equations in axisymmetric tokamak geometry \cite{Brizard_Tronko}.

\subsection{Guiding-center magnetization}

Using Eq.~\eqref{eq:mu_epsilon}, the lowest-order intrinsic guiding-center magnetic-dipole moment
\begin{equation}
\vb{\mu}_{\rm gc}^{(0)} \;=\; \frac{e}{2c}\;\left\langle\vb{\rho}_{0}\btimes \left( \Omega\;\pd{\vb{\rho}_{0}}{\zeta} \right)\right\rangle \;=\; 
-\,\mu\;\bhat,
\label{eq:mu_gc}
\end{equation}
yields the standard guiding-center magnetization \cite{Qin,Hazeltine_Meiss}
\begin{equation}
{\bf M}_{\rm gc}^{(0)} \;\equiv\; -\;\left(\int\;F\;\mu\;d^{3}P\right)\;\bhat \;\equiv\; -\;\frac{P_{\bot}\,\bhat}{B},
\label{eq:M_gc_0}
\end{equation}
which is expressed in terms of the guiding-center perpendicular pressure $P_{\bot}$. We note that Eq.~\eqref{eq:mu_gc} can also be obtained from the guiding-center Hamiltonian $H_{\rm gc} = p_{\|}^{2}/2m + \mu\,B$ as $\vb{\mu}_{\rm gc}^{(0)} \equiv -\,\partial H_{\rm gc}/\partial{\bf B} = -\,\mu\,
\bhat$. 

Lastly, we note that, up to first order in magnetic-field nonuniformity, the guiding-center representation of the plasma current density \eqref{eq:J_def} is
\begin{eqnarray}
{\bf J} & \simeq & {\bf J}_{\rm gc} \;+\; c\,\nabla\btimes{\bf M}_{\rm gc} \nonumber \\
 & = & \int \left[ e\,\frac{d_{\rm gc}{\bf X}}{dt}\;F \;-\; c\,\nabla\btimes\left(\mu\,F\;\bhat\right) \right]\;d^{3}P,
\end{eqnarray}
where the second-order guiding-center polarization current density $\partial{\bf P}_{\rm gc}/\partial t$, defined by Eq.~\eqref{eq:pol_current}, is omitted here.

\section{\label{sec:gy_pol}Linear and Nonlinear Gyrocenter Polarizations}

We now consider the polarization corrections to the guiding-center polarization density \eqref{eq:pi_gc} associated with the gyrocenter transformation 
${\cal T}_{\rm gy}$ from guiding-center phase-space coordinates $Z^{\alpha} = ({\bf X},p_{\|},\mu,\zeta)$ to gyrocenter phase-space coordinates 
$\ov{Z}^{\alpha} = (\ov{\bf X},\ov{p}_{\|},\ov{\mu},\ov{\zeta})$. Here, the total gyrocenter displacement from the particle position
\begin{eqnarray}
\ov{\vb{\rho}}_{\rm gy} & \equiv & {\sf T}_{\rm gy}^{-1}\left({\sf T}_{\rm gc}^{-1}{\bf x}\right) - \ov{\bf x} \;=\; {\sf T}_{\rm gy}^{-1}\left({\bf X} 
\;+\frac{}{} \vb{\rho}_{\rm gc}\right) - \ov{\bf X} \nonumber \\
 & = & \ov{\vb{\rho}}_{\rm gc} - \epsilon_{\delta}\,\left\{ \ov{S}_{1},\frac{}{} \ov{\bf X} + \ov{\vb{\rho}}_{\rm gc}\right\}_{\rm gc} - \epsilon_{\delta}^{2}\,\left\{ 
\ov{S}_{2},\frac{}{} \ov{\bf X} + \ov{\vb{\rho}}_{\rm gc}\right\}_{\rm gc} \nonumber \\
 &  &+\; \frac{\epsilon_{\delta}^{2}}{2}\;\left\{ \ov{S}_{1},\frac{}{} \{ \ov{S}_{1},\; \ov{\bf X} + \ov{\vb{\rho}}_{\rm gc}\}_{\rm gc} \right\}_{\rm gc} + \cdots
\label{eq:rho_gy_def}
\end{eqnarray}
includes the guiding-center displacement \eqref{eq:rho_gc} at the lowest order in $\epsilon_{\delta}$ (the dimensionless parameter associated with the perturbation amplitude), where $\ov{\vb{\rho}}_{\rm gc}$ denotes the guiding-center displacement \eqref{eq:rho_gc} evaluated in terms of the gyrocenter phase-space coordinates $\ov{Z}^{\alpha}$. 

The gyrocenter transformation from guiding-center phase space is generated by the canonical generating components $\ov{G}_{n}^{\alpha} \equiv \{ 
\ov{S}_{n},\; Z^{\alpha}\}_{\rm gc}$ ($n = 1, 2, ...)$, where the gyrocenter generating functions $\ov{S}_{1}$ and $\ov{S}_{2}$ satisfy
\begin{eqnarray}
\frac{d_{\rm gc}\ov{S}_{1}}{dt} & = & e\;\wt{\Phi}_{1{\rm gc}}, \label{eq:S1gy_eq} \\
\frac{d_{\rm gc}\ov{S}_{2}}{dt} & = & -\frac{e}{2}\left( \left\{ \ov{S}_{1},\frac{}{} \Phi_{1{\rm gc}}\right\}_{\rm gc} - \left\langle\left\{ \ov{S}_{1},\frac{}{} 
\Phi_{1{\rm gc}}\right\}_{\rm gc}\right\rangle\right) \nonumber \\
 &  &-\; e\;\left\{ \ov{S}_{1},\frac{}{} \langle\Phi_{1{\rm gc}}\rangle\right\}_{\rm gc}. \label{eq:S2gy_eq}
\end{eqnarray}
Here, the gyroangle average $\langle\cdots\rangle$ is associated with the fast-gyromotion averaging and $\wt{\Phi}_{1{\rm gc}} \equiv \Phi_{1{\rm gc}} - \langle\Phi_{1{\rm gc}}\rangle$ denotes the gyroangle-dependent part of $\Phi_{1{\rm gc}} \equiv \Phi_{1}(\ov{\bf X} + \ov{\vb{\rho}}_{\rm gc},t)$. Since the functions $\ov{S}_{n}$ are explicitly gyroangle-dependent, we use $d_{\rm gc}/dt \simeq \Omega\,\partial/\partial\ov{\zeta}$ in 
Eqs.~\eqref{eq:S1gy_eq}-\eqref{eq:S2gy_eq} and we henceforth use the lowest-order form of the gyrocenter Poisson bracket
\begin{equation}
\{ F,\; G \}_{\rm gc} \;\simeq\; \pd{F}{\ov{\zeta}}\;\pd{G}{\ov{J}} \;-\; \pd{F}{\ov{J}}\;\pd{G}{\ov{\zeta}},
\label{eq:gcPB_0}
\end{equation}
where the gyrocenter gyroangle $\ov{\zeta}$ is canonically-conjugate to the gyrocenter gyro-action $\ov{J} \equiv \ov{\mu}\,B/\Omega$.

\subsection{Higher-order gyrocenter Hamiltonian}

We now consider the linear (first-order) and nonlinear (second-order) gyrocenter polarizations that arise from the gyrocenter dynamical reduction (i.e., as a second step after the guiding-center reduction). Here, we restrict our attention to the electrostatic case and briefly summarize the work presented by Mishchenko and Brizard \cite{Mish_Bri}, where the gyrocenter Hamiltonian
\begin{eqnarray} 
H_{\rm gy} & = & H_{\rm gc} \;+\; \epsilon_{\delta}\;e\;\langle\Phi_{1{\rm gc}}\rangle \;-\; \epsilon_{\delta}^{2}\,\frac{e}{2}\;\left\langle\left\{ 
S_{1},\frac{}{} \Phi_{1{\rm gc}}\right\}_{\rm gc}\right\rangle \nonumber \\
 &  &+\;\epsilon_{\delta}^{3}\,e \left[ \frac{1}{2}\;\left\langle\left\{ S_{1},\frac{}{} \{ S_{1},\; \langle\Phi_{1{\rm gc}}\rangle\}_{\rm gc}
\right\}_{\rm gc}\right\rangle \right. \nonumber \\
 &  &\left.+\; \frac{1}{3}\;\left\langle\left\{ S_{1},\frac{}{} \{ S_{1},\; \wt{\Phi}_{1{\rm gc}}\}_{\rm gc}\right\}_{\rm gc}\right\rangle \right]
\label{eq:Hgy_3}
\end{eqnarray}
contains terms up to third order in perturbation amplitude; in the remainder of this Section, we omit the overbar to denote gyrocenter phase-space coordinates and functions. 

We note that the electric-dipole-moment contribution to the guiding-center polarization \eqref{eq:P_gc_final} is obtained from the variation of the first-order gyrocenter Hamiltonian
\begin{eqnarray*}
H_{1{\rm gy}} \;=\; e\;\langle\Phi_{1{\rm gc}}\rangle & = & e\,\Phi_{1} \;-\; e\,\langle\vb{\rho}_{\rm gc}\rangle\bdot{\bf E}_{1} \\
 &  &-\; \frac{e}{2}\,\langle\vb{\rho}_{\rm gc}\,\vb{\rho}_{\rm gc}\rangle\bdot\nabla{\bf E}_{1} \;+\; \cdots,
\end{eqnarray*}
so that Eq.~\eqref{eq:pi_epsilon_var} yields
\begin{eqnarray}
\vb{\pi}_{\rm gc} & \equiv & -\,\pd{H_{1{\rm gy}}}{{\bf E}_{1}} \;+\; \nabla\bdot\left(\pd{H_{1{\rm gy}}}{\nabla{\bf E}_{1}} \right) \;+\; \cdots 
\nonumber \\
 & = & e\,\langle\vb{\rho}_{\rm gc}\rangle \;-\; \nabla\bdot\left(\frac{e}{2}\,\langle\vb{\rho}_{\rm gc}\,\vb{\rho}_{\rm gc}\rangle\right) \;+\; \cdots,
\label{eq:pi_gc_gy}
\end{eqnarray}
which gives Eq.~\eqref{eq:pi_gc}.

\subsection{Linear gyrocenter polarization}

If we consider the first-order term in Eq.~\eqref{eq:rho_gy_def}, we easily obtain the first-order gyrocenter polarization density
\begin{eqnarray} 
\vb{\pi}_{1{\rm gy}} & \equiv & e\;\langle\vb{\rho}_{1{\rm gy}}\rangle \;-\; \nabla\bdot\left( \frac{e}{2}\,\left\langle\vb{\rho}_{\rm gc}\,
\vb{\rho}_{1{\rm gy}} \;+\frac{}{} \vb{\rho}_{1{\rm gy}}\,\vb{\rho}_{\rm gc}\right\rangle\right) \nonumber \\
 & \simeq & -\,\frac{e^{2}}{\Omega}\,\pd{}{J}\left\langle \vb{\rho}_{0}\;\wt{\Phi}_{1{\rm gc}} - \nabla\bdot\left( 
\frac{\vb{\rho}_{0}\vb{\rho}_{0}}{2}\;\wt{\Phi}_{1{\rm gc}}\right)\right\rangle, 
\label{eq:rho1_gy_def}
\end{eqnarray}
where we used the lowest-order guiding-center Poisson bracket \eqref{eq:gcPB_0}, we use the lowest-order expression $\vb{\rho}_{\rm gc} \simeq 
\vb{\rho}_{0}$ for the guiding-center displacement \eqref{eq:rho_gc}, and we have included the electric-quadrupole-moment correction in 
Eq.~\eqref{eq:rho1_gy_def}. Here, the gyroangle-averaged first-order gyrocenter displacement is
\begin{eqnarray} 
\langle\vb{\rho}_{1{\rm gy}}\rangle & \simeq & -\;\frac{e}{\Omega}\;\pd{}{J}\left\langle \vb{\rho}_{0}\;\wt{\Phi}_{1{\rm gc}}\right\rangle \nonumber \\
 & = & -\,\frac{e}{m\Omega^{2}}\;\nabla_{\bot}\langle\Phi_{1{\rm gc}}\rangle \;\equiv\; \frac{\bhat}{\Omega}\btimes\frac{d_{1}{\bf X}}{dt},
\label{eq:rho1_gy}
\end{eqnarray}
where the perpendicular first-order gyrocenter drift-velocity
\begin{equation}
\left(\frac{d_{1}{\bf X}}{dt}\right)_{\bot} \;\equiv\; \frac{\bhat}{m\Omega}\btimes\nabla H_{1{\rm gy}} \;=\; \frac{c\bhat}{B}\btimes\nabla\langle
\Phi_{1{\rm gc}}\rangle
\label{eq:gy_velocity_1}
\end{equation} 
represents the perturbed $E \times B$ velocity. Hence, the electric-dipole-moment contribution to the linear gyrocenter polarization can be expressed in terms of the electric-dipole contribution (without the first-order FLR correction) 
\begin{equation}
{\bf P}_{\rm gy}^{(1)} \;=\; \int\;F\,\left( \frac{e\,\bhat}{\Omega}\btimes\frac{d_{1}{\bf X}}{dt}\right) d^{3}P,
\label{eq:P_gy_1}
\end{equation}
which appears naturally as the first-order gyrocenter correction to the guiding-center polarization \eqref{eq:P_gc_final}. 

According to the standard gyrokinetic (short-wavelength limit) ordering \cite{Brizard_Hahm}, Eq.~\eqref{eq:P_gy_1} yields a first-order gyrocenter polarization charge density $\varrho_{\rm pol\,(gy)}^{(1)} \equiv -\,\nabla\bdot{\bf P}_{\rm gy}^{(1)}$ and a second-order gyrocenter polarization current density 
${\bf J}_{\rm pol\,(gy)}^{(2)} \equiv \partial{\bf P}_{\rm gy}^{(1)}/\partial t$, which is of the same order as the guiding-center polarization current density derived from Eq.~\eqref{eq:P_gc_final}. In the long-wavelength limit, however, the first-order gyrocenter polarization charge density 
$-\,\nabla\bdot{\bf P}_{\rm gy}^{(1)}$ becomes a second-order effect that is comparable to the guiding-center polarization charge density.

In order to derive the simplest form for the nonlinear (second-order) gyrocenter polarization density, we will evaluate the first-order gyrocenter displacement \eqref{eq:rho1_gy} in the zero-Larmor-radius (ZLR) limit (indicated by $\rightarrow$). In this limit, the second-order gyrocenter Hamiltonian becomes
\begin{equation}
H_{2{\rm gy}} \;\simeq\; -\,\frac{e^{2}}{2\Omega}\;\pd{}{J} \left\langle \wt{\Phi}_{1{\rm gc}}^{2}\right\rangle \;\rightarrow\; -\;\frac{1}{2}\,m\,
\Omega^{2}\;|\vb{\xi}_{1{\rm gy}}|^{2},
\label{eq:H2gy_ZLR}
\end{equation}
the first-order gyrocenter displacement \eqref{eq:rho1_gy} becomes
\begin{equation}
\langle\vb{\rho}_{1{\rm gy}}\rangle \;\rightarrow\; \vb{\xi}_{1{\rm gy}} \;\equiv\; \frac{c\,{\bf E}_{1}}{B\Omega},
\label{eq:xi1_gy}
\end{equation}
and the electric-quadrupole-moment contribution in Eq.~\eqref{eq:rho1_gy_def} vanishes, so that the first-order gyrocenter polarization density is 
$\vb{\pi}_{1{\rm gy}} \equiv -\,\partial H_{2{\rm gy}}/\partial{\bf E}_{1} \rightarrow e\,\vb{\xi}_{1{\rm gy}}$.

\subsection{Nonlinear gyrocenter polarization}

We now derive the nonlinear (quadratic) gyrocenter polarization from the third-order gyrocenter Hamiltonian. For this purpose, we use the ZLR limit and the simplified gyrocenter Poisson bracket \eqref{eq:gcPB_0}. 

\subsubsection{Variational derivation}

With these approximations (see App.~\ref{sec:gy} for details), we obtain the cubic gyrocenter Hamiltonian \cite{Mish_Bri}
\begin{eqnarray}
H_{3{\rm gy}} & \simeq & \frac{e^{3}}{2\,\Omega^{2}}\;\pd{}{J} \left\langle \left(\wt{\Phi}_{1{\rm gc}}\right)^{2}\;\pd{\Phi_{1{\rm gc}}}{J}\right\rangle 
\nonumber \\
 & \rightarrow & -\,\frac{e}{2}\,{\bf E}_{1}\bdot\nabla\left(\frac{1}{2}\,|\vb{\xi}_{1{\rm gy}}|^{2}\right),
\label{eq:Hgy_3}
\end{eqnarray}
where we used the definition \eqref{eq:xi1_gy}. It is now clear that the third-order gyrocenter Hamiltonian \eqref{eq:Hgy_3} explicitly contains dipole and quadrupole contributions. The nonlinear gyrocenter polarization density is therefore expressed as
\begin{eqnarray}
\vb{\pi}_{2{\rm gy}} & \equiv & -\;\pd{H_{3{\rm gy}}}{{\bf E}_{1}} \;+\; \nabla\bdot\left(\pd{H_{3{\rm gy}}}{(\nabla{\bf E}_{1})} \right) \nonumber \\
 & \rightarrow & \frac{e}{2} \left( \nabla\vb{\xi}_{1{\rm gy}}\bdot\vb{\xi}_{1{\rm gy}} \;+\; \vb{\xi}_{1{\rm gy}}\bdot\nabla\vb{\xi}_{1{\rm gy}} \right) 
\nonumber \\
 &  &-\; \nabla\bdot\left(\frac{e}{2}\;\vb{\xi}_{1{\rm gy}}\,\vb{\xi}_{1{\rm gy}} \right) \nonumber \\
 & \simeq & e\;\nabla\vb{\xi}_{1{\rm gy}}\bdot\vb{\xi}_{1{\rm gy}} \;-\; \nabla\bdot\left(\frac{e}{2}\;\vb{\xi}_{1{\rm gy}}\,\vb{\xi}_{1{\rm gy}} \right),
\label{eq:pi2_gy_def}
\end{eqnarray}
where we used the fact that the first-order gyrocenter displacement \eqref{eq:xi1_gy} is nearly curl-free (i.e., $\nabla\vb{\xi}_{1{\rm gy}}\bdot
\vb{\xi}_{1{\rm gy}} \simeq \vb{\xi}_{1{\rm gy}}\bdot\nabla\vb{\xi}_{1{\rm gy}}$). 

The first term in Eq.~\eqref{eq:pi2_gy_def} comes from the electric-dipole contribution derived from the second-order gyrocenter Hamiltonian 
\eqref{eq:H2gy_ZLR}:
\begin{eqnarray} 
\vb{\pi}_{2{\rm gy}}^{\rm (dip)} & = & \frac{e\,\bhat}{\Omega}\btimes\frac{d_{2}{\bf X}}{dt} \;=\; 
\frac{e\,\bhat}{\Omega}\btimes\left( \frac{\bhat}{m\Omega}\btimes\nabla H_{2{\rm gy}} \right) \nonumber \\
 & \rightarrow & e\;\nabla_{\bot}\vb{\xi}_{1{\rm gy}}\bdot\vb{\xi}_{1{\rm gy}},
\label{eq:pi_2_dip}
\end{eqnarray}
which was considered by Lee and Kolesnikov \cite{Lee_Koles_1,Lee_Koles_2}. 

Lastly, we compare the nonlinear gyrocenter polarization density \eqref{eq:pi2_gy_def} with the oscillation-center polarization density
\eqref{eq:pi_oc}. For this purpose, we use the eikonal representation \eqref{eq:eikonal_EB} for $\wt{\vb{\xi}}_{1{\rm gy}} = \vb{\xi}_{1{\rm gy}}\;
\exp(i\,\epsilon^{-1}\Theta) + {\rm c.c.}$ and obtain the eikonal-averaged expression (denoted by a hat)
\begin{equation} 
\wh{\vb{\pi}}_{2{\rm gy}} \;=\; e\,{\bf k}_{\bot}\btimes\left(i\frac{}{}\vb{\xi}_{1{\rm gy}}\btimes\vb{\xi}_{1{\rm gy}}^{*}\right),
\label{eq:pi2_gy_eik}
\end{equation}
which is identical in form to the oscillation-center polarization density \eqref{eq:pi_oc} with an important difference. While the
oscillation-center displacement $\vb{\xi}_{1{\rm oc}} \propto m^{-1}$ produces a second-order polarization density $\vb{\pi}_{2{\rm oc}} 
\propto m^{-2}$ that favors the contribution of electrons, we note that the gyrocenter displacement $\vb{\xi}_{1{\rm gy}} \propto m$ produces a second-order polarization density $\wh{\vb{\pi}}_{2{\rm gy}} \propto m^{2}$ that favors the contribution of ions.

\subsubsection{Higher-order quasineutrality condition}

Next, we note that, when we apply the quasineutrality condition in gyrokinetic theory (in the long-wavelength limit), the gyrokinetic Poisson equation \eqref{eq:varrho_def} is replaced with
\begin{eqnarray}
n_{\rm e} & = & N_{\rm i} \;-\; \nabla\bdot\left[ N_{\rm i} \left(\epsilon_{\delta}\;\vb{\xi}_{1{\rm gy}} \;+\frac{}{} \epsilon_{\delta}^{2}\,
\vb{\xi}_{2{\rm gy}} \;+\; \cdots \right) \right] \nonumber \\
 &  &-\; \epsilon_{\rm B}^{2}\;\nabla\bdot\left(e^{-1}\;{\bf P}_{\rm gc}\right),
\label{eq:gyro_quasi}
\end{eqnarray}
where $n_{\rm e}$ denotes the electron particle density while $N_{\rm i}$ denotes the ion gyrocenter density \cite{Brizard_92}, and the second-order gyrocenter displacement is defined as
\begin{equation}
\vb{\xi}_{2{\rm gy}} \;\equiv\; \frac{1}{2}\left[ \vb{\xi}_{1{\rm gy}}\bdot\nabla\vb{\xi}_{1{\rm gy}} \;-\frac{}{} \vb{\xi}_{1{\rm gy}}\;\left(\nabla
\bdot\vb{\xi}_{1{\rm gy}}\right) \right].
\label{eq:xi_gy_2_def}
\end{equation}
As mentioned above, the guiding-center polarization charge density (last term on the right side), which is always ignored in standard gyrokinetic theory \cite{Brizard_Hahm}, is at least of the same order as the linear gyrocenter polarization charge density in the long-wavelength limit. 

It is precisely in this long-wavelength limit that Parra and Catto \cite{Parra_1,Parra_2} have argued that standard gyrokinetic theory 
\cite{Brizard_Hahm} is incomplete. By following a one-step iterative procedure (which combines the guiding-center and gyrocenter transformations), Parra and Catto obtain a long-wavelength quasineutrality equation [see Eq.~(55) of Ref.~\cite{Parra_1}] that ignores the terms associated with the nonlinear gyrocenter polarization and the guiding-center polarization in Eq.~\eqref{eq:gyro_quasi} and adds the nonlinear correction 
\begin{equation}
-\,\frac{N_{\rm i}}{2}\;\left( \frac{m_{\rm i}c^{2}}{T_{\rm i}B^{2}}\;|\nabla_{\bot}\Phi_{1}|^{2} \right) \;\equiv\; 
-\,\frac{N_{\rm i}}{2}\;\left(\frac{|\vb{\xi}_{1{\rm gy}}|}{\rho_{\rm th(i)}}\right)^{2},
\label{eq:Parra}
\end{equation}
to the ion gyrocenter density $N_{\rm i}$, where the ordering \cite{Dimits} 
\begin{equation}
(k_{\bot}\rho_{\rm th(i)})\,\frac{e\,\Phi_{1}}{T_{\rm i}} \;\equiv\; \frac{|\vb{\xi}_{1{\rm gy}}|}{\rho_{\rm th(i)}} \;\ll\; 1
\label{eq:Dimits}
\end{equation}
is applied. We remark that the Parra-Catto nonlinear correction \eqref{eq:Parra} is not a polarization-charge correction (since it is not expressed as a spatial divergence) and it lacks a proper cold-ion limit (i.e., it diverges as $T_{\rm i} \rightarrow 0$). In fact, it was shown by Mishchenko and Brizard \cite{Mish_Bri} that Eq.~\eqref{eq:Parra} is a fictitious correction that can be made to disappear as a result of a careful two-step approach to gyrokinetic theory; see the discussion surrounding Eqs.~(48)-(52) in Ref.~\cite{Mish_Bri} and the recent paper by Miyato, Scott, and Yagi \cite{BDS}.

\subsection{Previous nonlinear polarization densities}

We now compare this second-order gyrocenter polarization density \eqref{eq:pi2_gy_def} with previous results.

\subsubsection{Dubin {\it et al.} \cite{Dubin}}

The first systematic derivation of the gyrocenter polarization came from the pull-back expression $\langle \delta_{\rm gc}^{3}\,{\sf T}_{\rm gy}F\rangle$ used by Dubin {\it et al.} \cite{Dubin}, where $\delta_{\rm gc}^{3} \equiv \delta^{3}({\bf X} + \vb{\rho}_{\rm gc} - {\bf r})$. A more practical form is obtained by integration by parts and involves the push-forward expression (given here up to second order)
\begin{eqnarray}
\left\langle{\sf T}_{\rm gy}^{-1}\delta_{\rm gc}^{3}\right\rangle & = & \langle\delta_{\rm gc}^{3}\rangle \;-\; \epsilon_{\delta}\,\frac{e}{\Omega}
\pd{}{J}\left\langle \wt{\Phi}_{1{\rm gc}}\;\delta_{\rm gc}^{3}\right\rangle \label{eq:push_delta} \\
 &  &+\; \frac{\epsilon_{\delta}^{2}e^{2}}{2\Omega^{2}} \pd{}{J} \left[ \left\langle \left( \pd{\wt{\Phi}_{1{\rm gc}}^{2}}{J} \;-\; 
\pd{\langle\wt{\Phi}_{1{\rm gc}}^{2}\rangle}{J}\right)\;\delta_{\rm gc}^{3} \right\rangle \right. \nonumber \\
 &  &\left.+\; \left\langle \wt{\Phi}_{1{\rm gc}}^{2}\;\pd{\delta_{\rm gc}^{3}}{J}\right\rangle \;+\; \left\langle \wt{\Phi}_{1{\rm gc}}\,
\delta_{\rm gc}^{3}\right\rangle\;\pd{\langle\Phi_{1{\rm gc}}\rangle}{J} \right], \nonumber
\end{eqnarray}
where, in the ZLR limit, the first-order term yields
\[ -\,\frac{e}{m\Omega^{2}}\,\nabla\Phi_{1}\bdot\nabla\delta^{3} \;\equiv\; \vb{\xi}_{1{\rm gy}}\bdot\nabla\delta^{3}. \]
Integration by parts then yields the first-order gyrocenter term in Eq.~\eqref {eq:gyro_quasi}.

The second-order terms in Eq.~\eqref{eq:push_delta} are expressed in the ZLR limit as
\begin{eqnarray}
 &  & \frac{1}{4}\,|\vb{\xi}_{1{\rm gy}}|^{2}\;\nabla^{2}\delta^{3} \;+\; \frac{1}{2}\;\vb{\xi}_{1{\rm gy}}\bdot\nabla\delta^{3}\;(\nabla\bdot
\vb{\xi}_{1{\rm gy}}) \nonumber \\
 &  &+\; 2\;\left\langle \left(\vb{\xi}_{1{\rm gy}}\bdot{\sf a}_{1}\bdot\vb{\xi}_{1{\rm gy}}\right)\frac{}{}{\sf a}_{1}:\nabla\nabla\delta^{3} \right\rangle \nonumber \\
 &  &+\; 4\;\left\langle \left(\vb{\xi}_{1{\rm gy}}\bdot{\sf a}_{1}\bdot\nabla\delta^{3}\right)\frac{}{}{\sf a}_{1}:\nabla\vb{\xi}_{1{\rm gy}} \right\rangle,
\label{eq:push_delta_2}
\end{eqnarray}
where the dyadic tensor ${\sf a}_{1}$ is defined in Eq.~\eqref{eq:a_1}. By using the identity \eqref{eq:a1_id}, we obtain
\begin{eqnarray*}
 &  &2\;\left\langle \left(\vb{\xi}_{1{\rm gy}}\bdot{\sf a}_{1}\bdot\vb{\xi}_{1{\rm gy}}\right)\frac{}{}{\sf a}_{1}:\nabla\nabla\delta^{3} \right\rangle \\
& &\hspace*{0.3in}= \frac{1}{2}\,\vb{\xi}_{1{\rm gy}}\vb{\xi}_{1{\rm gy}}:\nabla\nabla\delta^{3} \;-\; \frac{1}{4}\;|\vb{\xi}_{1{\rm gy}}|^{2}\;\nabla^{2}\delta^{3}, \\
 &  &4\;\left\langle \left(\vb{\xi}_{1{\rm gy}}\bdot{\sf a}_{1}\bdot\nabla\delta^{3}\right)\frac{}{}{\sf a}_{1}:\nabla\vb{\xi}_{1{\rm gy}} \right\rangle \\ 
& &\hspace*{0.3in}= (\vb{\xi}_{1{\rm gy}}\bdot\nabla\vb{\xi}_{1{\rm gy}})\bdot\nabla\delta^{3} \;-\; \frac{1}{2}\;(\vb{\xi}_{1{\rm gy}}\bdot\nabla
\delta^{3})\;\nabla\bdot\vb{\xi}_{1{\rm gy}},
\end{eqnarray*}
so that Eq.~\eqref{eq:push_delta_2} becomes
\begin{equation}
\frac{1}{2}\left[ \vb{\xi}_{1{\rm gy}}\bdot\nabla\vb{\xi}_{1{\rm gy}} - \vb{\xi}_{1{\rm gy}}\,(\nabla\bdot\vb{\xi}_{1{\rm gy}})\right]
\bdot\nabla\delta^{3} \equiv \vb{\xi}_{2{\rm gy}}\bdot\nabla\delta^{3},
\label{eq:push_delta_2_final}
\end{equation}
where we used the definition \eqref{eq:xi_gy_2_def} and, upon integration by parts, we obtain the second-order gyrocenter term in 
Eq.~\eqref {eq:gyro_quasi}.

\subsubsection{Brizard \& Mishchenko \cite{Brizard_Mish}}

The gyrokinetic equations derived so far have been based on the ordering $e\,\Phi_{1}/T_{\rm i} \ll 1$, which is consistent with the ordering 
\eqref{eq:Dimits} in the short-wavelength (gyrokinetic) limit $k_{\bot}\rho_{\rm i} \sim 1$. In the long-wavelength limit $(k_{\bot}\rho_{\rm i} \ll 1)$, however, the ordering \eqref{eq:Dimits} yields $e\,\Phi_{1}/T_{\rm i} \sim 1$. With this modified ordering, Brizard and Mishchenko 
\cite{Brizard_Mish} derived the gyroangle-averaged gyrocenter displacement
\begin{equation}
\langle\vb{\rho}_{1{\rm gy}}\rangle \;\simeq\; \frac{\bhat}{\Omega}\btimes\left( \frac{c\bhat}{B}\btimes\nabla\Phi_{1} \;-\; \frac{c}{B\Omega}\;
\frac{d_{\rm gy}\nabla\Phi_{1}}{dt} \right),
\label{eq:rho_gy_BM}
\end{equation}
where $d_{\rm gy}/dt$ includes the $E\times B$ convective derivative and the second term in Eq.~\eqref{eq:rho_gy_BM} represents the effects of the polarization-drift velocity.

\section{\label{sec:sum}Summary}

In the present paper, we have presented two different approaches to deriving the reduced polarization associated with a near-identity transformation associated with the dynamical reduction of the Vlasov equation by Lie-transform perturbation methods. In the push-forward approach, the reduced polarization is constructed from the reduced displacement generated by the phase-space transformation. In the variational approach, the reduced polarization is constructed from derivatives of the reduced Hamiltonian with respect to the electric field and its gradients. Figure \ref{fig:Drift_pol} presented the geometry of the orbit-averaged reduced polarization shift, from which we obtain the reduced polarization.

We then proceeded to derive the reduced polarizations associated with the guiding-center transformation and the gyrocenter transformation. We thus obtain the reduced gyrocenter polarization (with quadrupole corrections)
\begin{eqnarray}
{\bf P}_{\rm gy} & \equiv & e\int\left[ F\;\langle\ov{\vb{\rho}}_{\rm gy}\rangle \,-\, \nabla\bdot\left( \frac{1}{2}\,F\;
\left\langle\ov{\vb{\rho}}_{\rm gy}\frac{}{}\ov{\vb{\rho}}_{\rm gy}\right\rangle \right)\right] d^{3}P \nonumber \\
 & = & \epsilon_{\rm B}\;{\bf P}_{\rm gc}^{(1)} \;+\; \epsilon_{\delta}\;{\bf P}_{\rm gy}^{(1)} \;+\; \epsilon_{\delta}^{2}\;{\bf P}_{\rm gy}^{(2)} + \cdots,
\label{eq:gy_pol_epsilon}
\end{eqnarray}
where the gyrocenter displacement \eqref{eq:rho_gy_def} includes effects due to the guiding-center and gyrocenter transformation. We also noted that the electric-dipole-moment contribution in Eq.~\eqref{eq:gy_pol_epsilon} is expressed in terms of the gyrocenter velocity:
\begin{equation}
{\bf P}_{\rm gy}^{{\rm (dip)}} \;=\; \frac{e\bhat}{\Omega}\btimes\left[\int\;F\;\left(\frac{d_{\rm gy}{\bf X}}{dt}\right)\,d^{3}P\right],
\label{eq:gy_pol_dip}
\end{equation}
which includes the (zeroth-order) guiding-center polarization (represented by the guiding-center velocity) as well as the higher-order linear and nonlinear gyrocenter polarizations (represented by the perturbed gyrocenter velocities). We noted that, in the long-wavelength limit, the guiding-center and linear gyrocenter polarization current densities are of the same order and, therefore, both should be kept. Furthermore, we noted that the nonlinear gyrocenter polarization has the same universal features of the oscillation-center polarization also derived by Lie-transformation perturbation method.

Lastly, we note that each additional dynamical reduction used in gyrokinetic theory (e.g., the bounce-center dynamical reduction \cite{Cary_Brizard}) introduces new polarization effects \cite{Wang_Hahm}.

\acknowledgments

This work was supported by a U.~S.~Dept.~of Energy grant under contract No.~DE-FG02-09ER55005.

\appendix

\section{\label{sec:gc}Guiding-center Displacement}

In this Appendix, we present the derivation of the guiding-center polarization density \eqref{eq:pi_gc} based on guiding-center Lie-transform perturbation theory \cite{RGL_83}. In order to recover the standard Pfirsch-Kaufman formula \cite{Pfirsch,Kaufman_86} given by Eq.~\eqref{eq:pi_gc}, however, the guiding-center transformation needs to be modified at first order in magnetic-field nonuniformity. As a result of this new transformation, the guiding-center phase-space Lagrangian is now
\begin{eqnarray}
\Gamma_{\rm gc} & = & \left( \frac{e}{c}\,{\bf A} \;+\; p_{\|}\,\bhat \;-\; \frac{1}{2}\,J\,\nabla\btimes\bhat\right)\bdot\exd{\bf X} \nonumber \\
 &  &+\; J\;\left(\exd\zeta \;-\frac{}{}{\bf R}\bdot\exd{\bf X}\right) \;-\; H_{\rm gc}\,\exd t,
\label{eq:gc_psl}
\end{eqnarray}
where $J \equiv \mu B/\Omega$ denotes the guiding-center gyroaction, ${\bf R}$ denotes the gyrogauge vector field \cite{Cary_Brizard}, and the term 
$\nabla\btimes\bhat \equiv \tau\,\bhat + \bhat\btimes\vb{\kappa}$ includes the standard magnetic-twist term $\tau \equiv \bhat\bdot\nabla\btimes\bhat$ and the new correction term involving magnetic curvature $\vb{\kappa} \equiv\bhat\bdot\nabla\bhat$. 

The guiding-center phase-space transformation leading to Eq.~\eqref{eq:gc_psl} involves the generating-vector-field components $G_{1}^{\bf x} = -\,\vb{\rho}_{0}$, which defines the lowest-order gyroradius, and
\begin{eqnarray}
G_{2}^{\bf x} & = & G_{2\|}^{\bf x}\;\bhat + \rho_{\|}\,\tau \;\vb{\rho}_{0} + \frac{1}{2} \left( G_{1}^{J} \;-\frac{}{} J\vb{\rho}_{0}\bdot\nabla\ln B 
\right) \pd{\vb{\rho}_{0}}{J} \nonumber \\
 &  &+\; \frac{1}{2} \left( G_{1}^{\zeta} \;+\frac{}{} \vb{\rho}_{0}\bdot{\bf R} \right) \pd{\vb{\rho}_{0}}{\zeta} \;+\; G_{2\,{\rm (pol)}}^{\bf x},
\label{eq:G2_x}
\end{eqnarray}
where the guiding-center polarization correction $G_{2\,{\rm (pol)}}^{\bf x}$ is determined below [see Eq.~\eqref{eq:G2_x_pol}], while
\begin{eqnarray}
G_{2\|}^{\bf x} & = & \frac{2\,p_{\|}}{m\Omega}\;\pd{\vb{\rho}_{0}}{\theta}\bdot\vb{\kappa} \;+\; \frac{J}{m\Omega}\;\left({\sf a}_{2}:\nabla\bhat\right),
\label{eq:G2_par} \\
G_{1}^{J} & = & \vb{\rho}_{0}\bdot\left( J\nabla\ln B \;+\; \frac{p_{\|}^{2}}{m\Omega}\;\vb{\kappa} \right) \nonumber \\
 &  &-\; \frac{J\,p_{\|}}{m\Omega}\; \left( \tau \;+\; {\sf a}_{1}:\nabla\bhat \right),
\label{eq:G1_J} \\
G_{1}^{\zeta} & = & -\;\vb{\rho}_{0}\bdot{\bf R} \;+\; \frac{p_{\|}}{m\Omega}\;\left( {\sf a}_{2}:\nabla\bhat \right) \nonumber \\
 &  &+\; \pd{\vb{\rho}_{0}}{\zeta}\bdot\left( \nabla\ln B \;+\; \frac{p_{\|}^{2}\,\vb{\kappa}}{2m\, J\Omega}\right),
\label{eq:G1_zeta}
\end{eqnarray}
where the dyadic tensor fields in Eqs.~\eqref{eq:G2_par}-\eqref{eq:G1_p} are defined as
\begin{eqnarray}
{\sf a}_{1} & \equiv & -\;\frac{1}{2}\left( \wh{\bot}\,\wh{\rho} \;+\frac{}{} \wh{\rho}\,\wh{\bot}\right) \;\equiv\; \pd{{\sf a}_{2}}{\zeta},
\label{eq:a_1} \\
{\sf a}_{2} & \equiv & \frac{1}{4}\left( \wh{\bot}\,\wh{\bot} \;-\frac{}{} \wh{\rho}\,\wh{\rho}\right) \;\equiv\; -\,\frac{1}{4}\;
\pd{{\sf a}_{1}}{\zeta}.
\label{eq:a_2}
\end{eqnarray}
Using the additional component
\begin{equation}
G_{1}^{p_{\|}} \;=\; -\;p_{\|}\,\vb{\rho}_{0}\bdot\vb{\kappa} \;+\; J\,\left(\tau \;+\frac{}{} {\sf a}_{1}:\nabla\bhat\right),
\label{eq:G1_p}
\end{equation}
we can show that the guiding-center Jacobian is
\begin{equation}
{\cal J}_{\rm gc} \;\equiv\; B \;-\; \epsilon_{B}\,\pd{}{Z^{\alpha}}\left(B\;G_{1}^{\alpha}\right) + \cdots \;=\; B_{\|}^{*},
\label{eq:Jac_gc}
\end{equation}
where $B_{\|}^{*} \equiv \bhat\bdot{\bf B}^{*}$ with 
\begin{eqnarray}
{\bf B}^{*} & = & {\bf B} \;+\; \epsilon_{B}\,\nabla\btimes\left(\frac{cp_{\|}}{e}\,\bhat\right) \nonumber \\
 & &-\; \epsilon_{B}^{2}\,\nabla\btimes\left(\frac{cJ}{e}\,{\bf R} + \frac{cJ}{2\,e}\,\,\nabla\btimes\bhat\right).
\end{eqnarray} 

The first-order guiding-center displacement is expressed from Eq.~\eqref{eq:rho_epsilon_def} as
\begin{eqnarray}
\vb{\rho}_{1{\rm gc}} & \equiv & -\;G_{2}^{\bf x} \;+\; \frac{1}{2}\,\vb{\rho}_{0}\bdot\nabla\vb{\rho}_{0} \nonumber \\
 &  &-\; \frac{1}{2} \left( G_{1}^{J}\;\pd{\vb{\rho}_{0}}{J} \;+\; G_{1}^{\zeta}\;\pd{\vb{\rho}_{0}}{\zeta} \right),
\label{eq:rho1_gc_def}
\end{eqnarray}
where 
\begin{eqnarray}
\vb{\rho}_{0}\bdot\nabla\vb{\rho}_{0} & = & -\,\left(J\vb{\rho}_{0}\bdot\nabla\ln B\right) \pd{\vb{\rho}_{0}}{J} \;-\; \left(\vb{\rho}_{0}\bdot{\bf R}
\right) \pd{\vb{\rho}_{0}}{\zeta} \nonumber \\
 &  &-\; \left(\vb{\rho}_{0}\bdot\nabla\bhat\bdot\vb{\rho}_{0}\right)\bhat.
\label{eq:rhogradrho}
\end{eqnarray}
By combining Eqs.~\eqref{eq:G2_x}-\eqref{eq:G1_zeta} and Eq.~\eqref{eq:rhogradrho} into the first-order guiding-center displacement 
\eqref{eq:rho1_gc_def}, we find
\begin{eqnarray}
\vb{\rho}_{1{\rm gc}} & = & -\,\left( G_{2\|}^{\bf x} \;+\; \frac{1}{2}\,\vb{\rho}_{0}\bdot\nabla\bhat\bdot\vb{\rho}_{0}\right)\;\bhat \;-\; \rho_{\|}\tau\;\vb{\rho}_{0} \label{eq:rho1_gc} \\
 &  &-\; G_{1}^{J}\;\pd{\vb{\rho}_{0}}{J} \;-\; \left( G_{1}^{\zeta} \;+\frac{}{} \vb{\rho}_{0}\bdot{\bf R}\right)\;\pd{\vb{\rho}_{0}}{\zeta} \;-\; 
G_{2\,{\rm (pol)}}^{\bf x}.
\nonumber
\end{eqnarray}
The gyroangle-averaged guiding-center displacement, thus, becomes
\begin{eqnarray}
\langle\vb{\rho}_{1{\rm gc}}\rangle & = & -\,\frac{J}{2m\Omega}\,\nabla\bdot\left(\bhat\,\bhat\right) \;+\; \left(
\frac{J\,\vb{\kappa}}{2\,m\Omega} \;-\; G_{2\,{\rm (pol)}}^{\bf x} \right) \nonumber \\
 &  &-\; \left\langle G_{1}^{J}\;\pd{\vb{\rho}_{0}}{J}\right\rangle  \;-\; \left\langle\left( G_{1}^{\zeta} \;+\frac{}{} \vb{\rho}_{0}\bdot{\bf R}\right)\pd{\vb{\rho}_{0}}{\zeta}\right\rangle \nonumber \\
 & = & \frac{\bhat}{\Omega}\btimes\frac{d_{\rm gc}{\bf X}}{dt} \;+\; \nabla\bdot\left(\frac{1}{2}\,\left\langle\vb{\rho}_{0}\frac{}{}\vb{\rho}_{0}\right\rangle\right) \nonumber \\
 &  &+\; \left(\frac{J\,\vb{\kappa}}{2\,m\Omega} \;-\; G_{2\,{\rm (pol)}}^{\bf x} \right).
\end{eqnarray}
Hence, we obtain the standard Pfirsch-Kaufman result \cite{Pfirsch,Pfirsch_Morrison,Kaufman_86} for the guiding-center polarization density 
\begin{eqnarray}
\vb{\pi}_{\rm gc} & \equiv & e\,\langle\vb{\rho}_{1{\rm gc}}\rangle \;-\; \nabla\bdot\left(\frac{e}{2}\,\left\langle\vb{\rho}_{0}\frac{}{}\vb{\rho}_{0}\right\rangle\right) \nonumber \\
 & = & \frac{e\,\bhat}{\Omega}\btimes\frac{d_{\rm gc}{\bf X}}{dt},
\label{eq:pi_gc_Lie}
\end{eqnarray}
if the guiding-center polarization correction is
\begin{equation}
G_{2\,{\rm (pol)}}^{\bf x} \;\equiv\; \frac{J\,\vb{\kappa}}{2\,m\Omega} \;=\; \frac{\bhat}{\Omega}\btimes\left[\nabla\btimes\left( -\,
\frac{J\,\bhat}{2\,m}\right)\right].
\label{eq:G2_x_pol}
\end{equation} 

\section{\label{sec:gy}Third-order Gyrocenter Hamiltonian}

In this Appendix, we use the lowest-order guiding-center displacement $\vb{\rho}_{\rm gc} = \vb{\rho}_{0}$ to calculate the lowest-order cubic gyrocenter Hamiltonian \cite{Mish_Bri}
\begin{eqnarray}
H_{3{\rm gy}} & = & \frac{e}{2}\;\left\langle\left\{ \ov{S}_{1},\frac{}{} \{ \ov{S}_{1},\; \langle\Phi_{1{\rm gc}}\rangle\}_{\rm gc}
\right\}_{\rm gc}\right\rangle \nonumber \\
 &  &+\; \frac{e}{3}\;\left\langle\left\{ \ov{S}_{1},\frac{}{} \{ \ov{S}_{1},\; \wt{\Phi}_{1{\rm gc}}\}_{\rm gc}\right\}_{\rm gc}\right\rangle,
\label{eq:H3gy_def}
\end{eqnarray}
where we use the lowest-order guiding-center Poisson bracket to find
\begin{eqnarray*}
\left\langle\left\{ \ov{S}_{1},\frac{}{} \{ \ov{S}_{1},\; \langle\Phi_{1{\rm gc}}\rangle\}_{\rm gc}\right\}_{\rm gc}\right\rangle & \simeq & \pd{}{J}\left(
\left\langle\wt{\Phi}_{1{\rm gc}}^{2}\right\rangle\;\pd{\langle\Phi_{1{\rm gc}}\rangle}{J}\right), \\
\left\langle\left\{ \ov{S}_{1},\frac{}{} \{ \ov{S}_{1},\; \wt{\Phi}_{1{\rm gc}}\}_{\rm gc}\right\}_{\rm gc}\right\rangle & \simeq & \frac{3}{2}\;\pd{}{J}
\left\langle \wt{\Phi}_{1{\rm gc}}^{2}\;\pd{\wt{\Phi}_{1{\rm gc}}}{J}\right\rangle,
\end{eqnarray*}
so that Eq.~\eqref{eq:H3gy_def} becomes
\begin{equation}
H_{3{\rm gy}} \;\simeq\; \frac{e^{3}}{2\,\Omega^{2}}\;\pd{}{J} \left\langle \left(\wt{\Phi}_{1{\rm gc}}\right)^{2}\;
\pd{\Phi_{1{\rm gc}}}{J}\right\rangle.
\label{eq:H3_gy}
\end{equation}
Next, the Taylor-expansion of $\Phi_{1{\rm gc}}$ in powers of $\vb{\rho}_{0}$ up to second order yields
\begin{eqnarray}
\Phi_{1{\rm gc}} & = & \Phi_{1} \;+\; \vb{\rho}_{0}\bdot\nabla\Phi_{1} \nonumber \\
 &  &+\; \frac{J}{2m\Omega} \left( \nabla_{\bot}^{2}\Phi_{1} \;-\frac{}{} 4\,{\sf a}_{2}:\nabla\nabla\Phi_{1} \right),
\end{eqnarray}
so that we find
\begin{eqnarray}
\left\langle \wt{\Phi}_{1{\rm gc}}^{2}\;\pd{\Phi_{1{\rm gc}}}{J}\right\rangle & = & \frac{J}{2\;(m\Omega)^{2}}\;|\nabla_{\bot}\Phi_{1}|^{2}\,
\nabla_{\bot}^{2}\Phi_{1} \nonumber \\
 & + &\frac{4\,J}{(m\Omega)^{2}}\;\left\langle \nabla\Phi_{1}\bdot{\sf a}_{1}\bdot\nabla\Phi_{1}\frac{}{} {\sf a}_{1}:\nabla
\nabla\Phi_{1}\right\rangle,
\nonumber 
\end{eqnarray}
where we used ${\sf a}_{1} \equiv \partial{\sf a}_{2}/\partial\zeta$. By using the identity
\begin{eqnarray}
 &  &4\,\left\langle (\nabla\psi\bdot{\sf a}_{1}\bdot\nabla\chi)\frac{}{} {\sf a}_{1}:\nabla\nabla\phi\right\rangle \nonumber \\
 & &\hspace*{1in}\equiv\; \nabla_{\bot}\psi\bdot\nabla\,\nabla\phi\bdot\nabla_{\bot}\chi \nonumber \\
 &  &\hspace*{1.2in}-\; \frac{1}{2}(\nabla_{\bot}\psi\bdot\nabla_{\bot}\chi)\;\nabla_{\bot}^{2}\phi,
\label{eq:a1_id} 
\end{eqnarray}
where $(\psi,\chi,\phi)$ are arbitrary functions, Eq.~\eqref{eq:H3_gy} yields the long-wavelength result
\begin{eqnarray}
H_{3{\rm gy}} & \rightarrow & \frac{e^{3}}{2\,(m\Omega^{2})^{2}} \left( \frac{1}{2}\,|\nabla\Phi_{1}|^{2}\;\nabla^{2}\Phi_{1} \right. \nonumber \\
 &  &\left.+\; \nabla\Phi_{1}\nabla\Phi_{1}:\nabla\nabla\Phi_{1} \;-\; \frac{1}{2}\,|\nabla\Phi_{1}|^{2}\;\nabla^{2}\Phi_{1} \right) \nonumber \\
 & = & -\,\frac{e}{2}\;{\bf E}_{1}\bdot\nabla\left(\frac{1}{2}\,|\vb{\xi}_{1{\rm gy}}|^{2}\right).
\label{eq:H3gy_ZLR}
\end{eqnarray}

\end{document}